\documentclass{wscpaperproc}
\usepackage{latexsym}
\usepackage{graphicx}
\usepackage{mathptmx}
\usepackage[T1]{fontenc}
\usepackage{subcaption}

\usepackage{amsmath}
\usepackage{amsfonts}
\usepackage{amssymb}
\usepackage{amsbsy}
\usepackage{amsthm}

\usepackage[pdftex,colorlinks=true,urlcolor=blue,citecolor=black,anchorcolor=black,linkcolor=black]{hyperref}

\newtheoremstyle{wsc}
{3pt}
{3pt}
{}
{}
{\bf}
{}
{.5em}
{}

\theoremstyle{wsc}

\begin{document}

%
%

\pagestyle{fancyplain}

\thispagestyle{plain}
\firstPageHead{}

\chead{\fancyplain{}{\itshape Farr, Talty, Farr, Stockdale, Cruickshank, and West}}

\rhead{}
\cfoot{}
\renewcommand{\headrulewidth}{0pt} 

\makeatletter
\let\@internalcite\cite
\def\cite{\def\@citeseppen{-1000}%
    \def\@cite##1##2{(##1\if@tempswa , ##2\fi)}%
    \def\citeauthoryear##1##2##3{##1 ##3}\@internalcite}
\def\citeNP{\def\@citeseppen{-1000}%
    \def\@cite##1##2{##1\if@tempswa , ##2\fi}%
    \def\citeauthoryear##1##2##3{##1 ##3}\@internalcite}
\def\citeN{\def\@citeseppen{-1000}%
    \def\@cite##1##2{##1\if@tempswa, ##2)\else{}\fi}%
    \def\citeauthoryear##1##2##3{##1 (##3)}\@citedata}
\def\citeA{\def\@citeseppen{-1000}%
    \def\@cite##1##2{(##1\if@tempswa , ##2\fi)}%
    \def\citeauthoryear##1##2##3{##1}\@internalcite}
\def\citeANP{\def\@citeseppen{-1000}%
    \def\@cite##1##2{##1\if@tempswa , ##2\fi}%
    \def\citeauthoryear##1##2##3{##1}\@internalcite}
\def\shortcite{\def\@citeseppen{-1000}%
    \def\@cite##1##2{(##1\if@tempswa , ##2\fi)}%
    \def\citeauthoryear##1##2##3{##2 ##3}\@internalcite}
\def\shortciteNP{\def\@citeseppen{-1000}%
    \def\@cite##1##2{##1\if@tempswa , ##2\fi}%
    \def\citeauthoryear##1##2##3{##2 ##3}\@internalcite}
\def\shortciteN{\def\@citeseppen{-1000}%
    \def\@cite##1##2{##1\if@tempswa, ##2\else{}\fi}%
    \def\citeauthoryear##1##2##3{##2 (##3)}\@citedata}
\def\shortciteA{\def\@citeseppen{-1000}%
    \def\@cite##1##2{(##1\if@tempswa , ##2\fi)}%
    \def\citeauthoryear##1##2##3{##2}\@internalcite}
\def\shortciteANP{\def\@citeseppen{-1000}%
    \def\@cite##1##2{##1\if@tempswa , ##2\fi}%
    \def\citeauthoryear##1##2##3{##2}\@internalcite}
\def\citeyear{\def\@citeseppen{-1000}%
    \def\@cite##1##2{(##1\if@tempswa , ##2\fi)}%
    \def\citeauthoryear##1##2##3{##3}\@citedata}
\def\citeyearNP{\def\@citeseppen{-1000}%
    \def\@cite##1##2{##1\if@tempswa , ##2\fi}%
    \def\citeauthoryear##1##2##3{##3}\@citedata}
%
%
%
\def\@citedata{%
    \@ifnextchar [{\@tempswatrue\@citedatax}%
                  {\@tempswafalse\@citedatax[]}%
}

\def\@citedatax[#1]#2{%
\if@filesw\immediate\write\@auxout{\string\citation{#2}}\fi%
  \def\@citea{}\@cite{\@for\@citeb:=#2\do%
    {\@citea\def\@citea{, }\@ifundefined
       {b@\@citeb}{{\bf ?}%
       \@warning{Citation `\@citeb' on page \thepage \space undefined}}%
{\csname b@\@citeb\endcsname}}}{#1}}%

%
\def\@citex[#1]#2{%
\if@filesw\immediate\write\@auxout{\string\citation{#2}}\fi%
  \def\@citea{}\@cite{\@for\@citeb:=#2\do%
    {\@citea\def\@citea{; }\@ifundefined
       {b@\@citeb}{{\bf ?}%
       \@warning{Citation `\@citeb' on page \thepage \space undefined}}%
{\csname b@\@citeb\endcsname}}}{#1}}%

%
\def\@biblabel#1{}
\makeatother



\newdimen\bibindent
\bibindent=0.0em
\def\thebibliography#1{\section*{\refname}\list
   {}{\settowidth\labelwidth{[#1]}
   \leftmargin\parindent
   \itemindent -\parindent
   \listparindent \itemindent
   \itemsep 0pt
   \parsep 0pt}
   \def\newblock{}
   \sloppy
   \sfcode`\.=1000\relax}


\setlength{\baselineskip}{12.7pt}

\title{Expert-in-the-Loop Systems with Cross-Domain and In-Domain Few-Shot Learning for Software Vulnerability Detection}

\author{\begin{center}David Farr\textsuperscript{1}, Kevin Talty\textsuperscript{2}, Alexandra Farr\textsuperscript{3}, John Stockdale\textsuperscript{2}, Iain Cruickshank\textsuperscript{4}, Jevin West\textsuperscript{1}\\
[11pt]
\textsuperscript{1}University of Washington, Seattle, WA, USA\\
\textsuperscript{2}United States Army\\
\textsuperscript{3}Microsoft, Redmond, WA, USA\\
\textsuperscript{4}Carnegie Mellon University, Pittsburgh, PA, USA\\
\end{center}
}

\maketitle

\vspace{-12pt}

\section*{Abstract}

As cyber threats become more sophisticated, rapid and accurate vulnerability detection is essential for maintaining secure systems. This study explores the use of Large Language Models (LLMs) in software vulnerability assessment by simulating the identification of Python code with known Common Weakness Enumerations (CWEs), comparing zero-shot, few-shot cross-domain, and few-shot in-domain prompting strategies. Our results indicate that while zero-shot prompting performs poorly, few-shot prompting significantly enhances classification performance, particularly when integrated with confidence-based routing strategies that improve efficiency by directing human experts to cases where model uncertainty is high, optimizing the balance between automation and expert oversight.

We find that LLMs can effectively generalize across vulnerability categories with minimal examples, suggesting their potential as scalable, adaptable cybersecurity tools in simulated environments. However, challenges such as model reliability, interpretability, and adversarial robustness remain critical areas for future research. By integrating AI-driven approaches with expert-in-the-loop (EITL) decision-making, this work highlights a pathway toward more efficient and responsive cybersecurity workflows. Our findings provide a foundation for deploying AI-assisted vulnerability detection systems in both real and simulated environments that enhance operational resilience while reducing the burden on human analysts.

\begin{figure*}[h!]
    \centering
    \includegraphics[scale=.48]{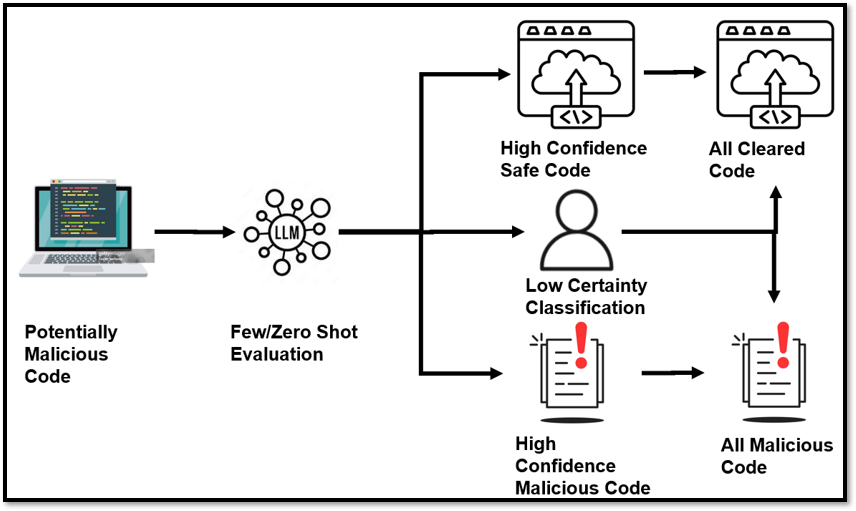}
    \caption{Incident response simulation of evaluating malicious software using generative artificial intelligence and uncertainty quantification for software routing.}
    \label{fig:sys_diagram}
\end{figure*}

\section{INTRODUCTION}
\label{sec:intro}

Cybersecurity and incident response have become increasingly critical to national security as governments, industries, and critical infrastructure remain dependent on software systems that are often vulnerable by design or deployment. The expanding attack surface, driven by rapid software development, interconnected systems, and persistent reliance on legacy technologies, poses significant challenges to securing essential infrastructure against evolving threats.

These challenges are further compounded by the accelerating pace of software production, fueled by generative artificial intelligence and the shift toward modular architectures that integrate containerized software from multiple development teams. Although modularity and containerization enhance flexibility and scalability, they also introduce substantial security risks. The heterogeneous nature of these ecosystems, where different components rely on diverse codebases, development methodologies, and dependencies, compromises security visibility and makes comprehensive vulnerability assessments increasingly difficult. As a result, undetected threats can persist within critical systems, potentially leading to severe breaches \cite{WEF}. Traditional methods of deploying fine-tuned models do not extrapolate well to novel vulnerabilities, require significant investments in data labeling, and do not have the ability for injection of context from subject matter experts.

To ensure safe software deployment and maintain trust in our systems, cybersecurity professionals need tools that can match the speed and complexity of modern software development. Large Language Models (LLMs) offer a compelling solution by augmenting vulnerability detection, automating security assessments, and identifying potential threats in real time. Their ability to process vast amounts of security-related data, detect patterns, and adapt to new threat landscapes makes them an invaluable asset in incident response workflows.

In this work, we simulate an incident response scenario in which a compromised environment results from the deployment of vulnerable code. We demonstrate how integrating LLMs with zero- and few-shot classification capabilities, enhanced by an uncertainty quantification metric, can significantly improve response effectiveness. By leveraging this approach, we enable a more robust analyst-in-the-loop evaluation, moving beyond standard incident response procedures that rely on broad, time-consuming queries to sift through terabytes of security logs. Our findings highlight the advantages of LLM-assisted triage and threat identification, ultimately contributing to a more adaptive and resilient cybersecurity framework for national security applications.

\section{Related Works}
Vulnerable software detection has long been a focus in machine learning research. Traditionally, classification models relied on large amounts of training data to develop deep learning models, yet these models often fail to generalize effectively to new or unseen vulnerabilities \cite{chen2023diversevul}. This challenge of poor generalization is well-documented across fine-tuned model applications in various domains \cite{ng2022my}. Notably, \cite{chen2023diversevul} demonstrate that LLMs outperform Graph Neural Networks (GNNs) that rely on code-structure representations for vulnerability detection.

This raises the critical question of how LLMs can be effectively integrated into larger incident response workflows using zero- and few-shot classification strategies. Zero-shot prompting leverages an LLM's base knowledge to classify vulnerabilities without requiring labeled examples, whereas few-shot prompting improves classification by incorporating example-based contextualization \cite{few}. Foundational LLMs have demonstrated strong performance in zero- and few-shot settings across multiple data classification tasks \cite{kojima2022large}\cite{ziems2024can}. 

Previous research (e.g., \cite{tamberg2025harnessing} and \cite{zhou2024large}) has explored the application of LLMs in software vulnerability detection under zero-, one-, and few-shot evaluation paradigms. \cite{zibaeirad2025reasoningllmszeroshotvulnerability} show that structured reasoning prompts significantly enhance LLM performance in vulnerability detection, with effectiveness varying by vulnerability type. \cite{ZHENG2024103992} introduce Few-VulD, a framework leveraging few-shot learning for software vulnerability detection, emphasizing the difficulties in training fine-tuned models due to the scarcity of labeled data across diverse software ecosystems. Additionally, \cite{chan2023transformerbasedvulnerabilitydetectioncode} demonstrates that LLMs can be integrated during software development and editing stages to reduce the effort required for post-deployment vulnerability fixes.

While these studies establish foundational approaches for using LLMs in software vulnerability detection and incident response, they largely overlook the role of human analysts in the security evaluation process. In real-world scenarios, human expertise remains critical for verifying and mitigating threats before software is redeployed in operational environments. \cite{googlepaper} demonstrated that LLMs can be used to reduce the workload of cybersecurity personnel by automating code vulnerability fixes. We build on this insight by optimizing the order of code evaluation by cybersecurity personnel after automated corrections.

Although little research has explored uncertainty quantification in software vulnerability detection for evaluation by analysts, significant progress has been made in other NLP applications. Throughout this paper, we refer to this human expert analyst as the expert-in-the-loop (EITL). \cite{tian-etal-2023-just} show that prompting language models for uncertainty can serve as an effective calibration technique. \cite{human} demonstrate that an external verifier using supervised models can effectively identify misclassified data. Furthermore, \cite{2024arXiv241013047F} show that leveraging the distance between top token log probabilities serves as an effective uncertainty quantification metric in constrained labeled spaces, requiring only a single model or API call. This approach reduces costs and increases efficiency for real-time incident response scenarios by ranking and prioritizing agent-driven vulnerability detection by uncertainty.

\vspace{5mm}

\section{Methodology}

Our methodology is illustrated in Figure \ref{fig:sys_diagram}. We simulate the evaluation of 1,096 Python software functions from the DIVERSEVUL dataset \cite{chen2023diversevul}, with approximately 10\% containing vulnerabilities classified under 37 known Common Weakness Enumerations (CWEs) \cite{MITRE}. To determine the appropriate response for each function, we use the confidence score metric shown in \cite{2024arXiv241013047F} to route code into one of three categories: automatic quarantine, cleared for deployment, or human analyst evaluation. This approach optimizes resource allocation by prioritizing low-confidence classifications for human review while automating high-confidence classifications.

Additionally, we test our model under three different levels of assumed incident response information, which influence how prompts are structured. The first level assumes no prior information, resulting in a zero-shot prompt. This represents cases where incident responders have no initial leads. The second level assumes a moderate amount of information, represented by a few-shot, cross-domain prompt. In this case, the model is provided with examples of vulnerabilities covering about one-third of the known CWEs, allowing it to generalize across unseen vulnerabilities. The final level assumes a well-defined target vulnerability, where the model is prompted with in-domain examples corresponding to a specific security threat.

\begin{figure*}[h!]
    \centering
    \includegraphics[scale=0.75]{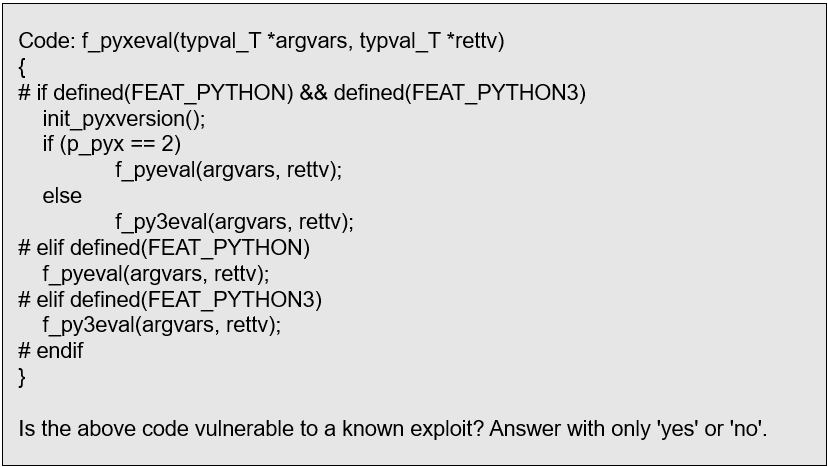}
    \caption{Example of a zero-shot prompt used for vulnerability classification.}
    \label{zero_prompt}
\end{figure*}
\subsection{Zero-Shot Prompt Design}

Our methodology is designed to be agnostic to prompt strategy, supporting strategies such as chain-of-thought (CoT), tree-of-thought (ToT), and other structured reasoning approaches. However, the model’s final output is constrained to a binary classification of whether code is vulnerable or not. This constraint ensures consistency in downstream processing and facilitates structured vulnerability routing.

For zero-shot prompting, we use a direct classification approach without additional reasoning steps. This simple strategy provides a baseline for performance evaluation and serves as a foundation for future refinements. An example zero-shot prompt used in a GPT API call is shown in Figure \ref{zero_prompt}.

\subsection{Few-Shot Cross-Domain Prompt Design}

For the few-shot cross-domain prompt, we provide the model with five examples of vulnerable code and five examples of non-vulnerable code. Then, task it with identifying vulnerable and non-vulnerable code. These examples are drawn from diverse vulnerability categories to encourage generalization beyond the provided samples. We did not experiment with providing incorrectly classified prompts. 

\begin{figure*}[h!]
    \centering
    \includegraphics[scale=0.75]{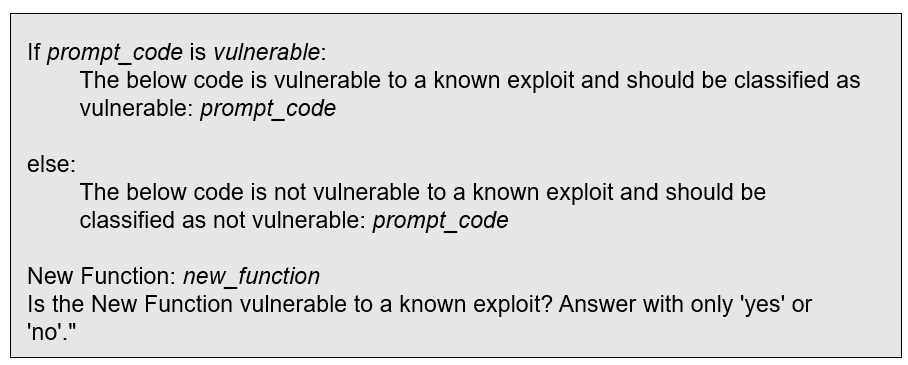}
    \caption{Example of a few-shot prompt used for vulnerability classification.}
    \label{few_prompt}
\end{figure*}

\subsection{Few-Shot In-Domain Prompt Design}

For the few-shot in-domain setting, we analyze seven CWEs that appear in "2024 CWE Top 25 Most Dangerous Software Weaknesses" \cite{MITRE} and our dataset of Python software functions. For each CWE type, we provide the model with one example of a vulnerable function and one example of a non-vulnerable function that fit under the CWE the model is attempting to classify. This scenario represents cases where incident responders have precise knowledge of the threat they are investigating. 

\subsection{Expert-in-the-Loop}

Fully automated incident response remains unlikely in the near future, making human oversight essential. To evaluate our methodology in an expert-in-the-loop setting, we compare two approaches: a baseline where all code is forwarded to an analyst without prior filtering and a certainty-based routing approach where only ambiguous cases are flagged for human review based on an uncertainty quantification metric. We made the necessary assumption that the expert-in-the-loop would always correctly classify the code it received. It is beyond the scope of this paper to evaluate when expert-in-the-loop classification accuracy is poor and how expert-in-the-loop classification is affected by model performance. The results from this section provide the upper limit of model performance when an expert-in-the-loop is constrained to only look at a subset of the code. 

\subsection{Certainty-Based Routing}

A core component of our methodology is certainty-based routing, which prioritizes human intervention for software functions where the model exhibits uncertainty. Analysts play a crucial role in verifying vulnerabilities, mitigating risks, and making final deployment decisions. To optimize analyst efficiency, we build upon the confidence metrics introduced by \cite{farr2024red}, using uncertainty quantification to prioritize ambiguous or high-risk cases.

The confidence score is defined as the absolute difference between the highest and second-highest token log probabilities within a constrained token set. Let \(\mathcal{T}\) represent the token set and \(P(t)\) denote the log probability of each token \(t \in \mathcal{T}\). The confidence score \(C\) is computed as:

\begin{equation}
    C = \left| \max_{t \in \mathcal{T}} P(t) - \max_{t \in \mathcal{T} \setminus \{t^*\}} P(t) \right|,
\label{csequation}
\end{equation}

where \(t^*\) is the token with the highest probability. A higher confidence score indicates greater certainty, while a lower confidence score suggests ambiguity, warranting human review.

To integrate certainty-based routing into incident response workflows, we define three classification thresholds based on the confidence score:

\begin{itemize}
    \item Automatic Quarantine: Code classified as vulnerable with high confidence are immediately flagged and quarantined for remediation.
    \item Cleared for Deployment: Code classified as non-vulnerable with high confidence are approved for deployment without further review.
    \item Human Analyst Evaluation: Code with intermediate confidence scores are flagged for human review.
\end{itemize}

\begin{figure*}[h!]
    \centering
    \includegraphics[scale=0.7]{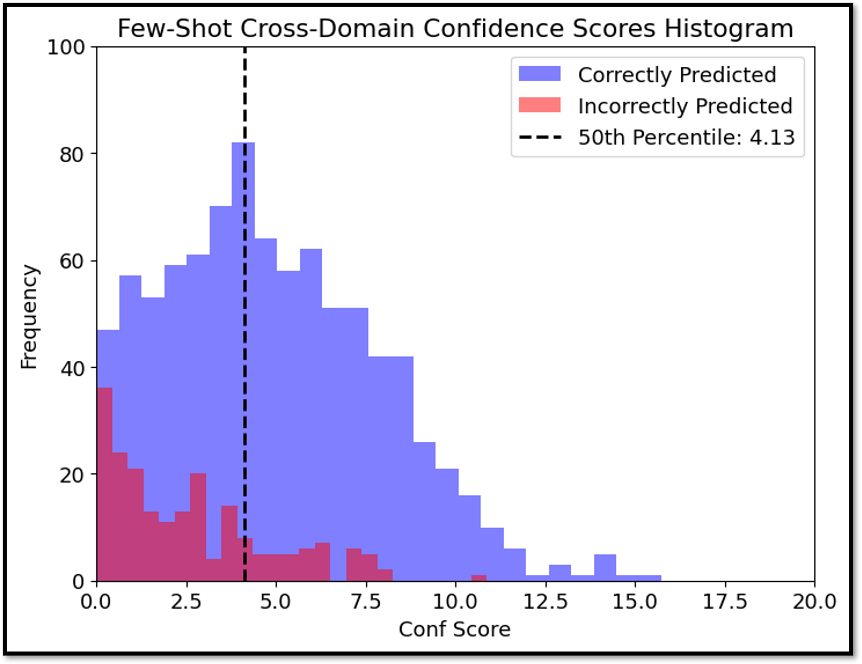}
    \caption{Example of confidence score distribution for the Few-Shot Cross-Domain strategy, with correct and incorrect classifications overlaid. In this scenario, where 50\% of the data is routed to an Expert-in-the-Loop (EITL), all instances to the left of the dashed line would be directed to the expert, capturing the majority of incorrect classifications for correction. Confidence scores, derived from differences in token log probabilities, commonly range between 10 and 20 using GPT-4o. However, due to GPT's underlying architecture, theoretical values can reach as high as 100.}
    \label{c_scores}
\end{figure*}

This routing strategy ensures that human analysts focus on cases where model uncertainty is high, effectively balancing automation with expert oversight to enhance detection accuracy and efficiency. Figure \ref{c_scores} illustrates this approach by depicting the distribution of confidence scores for correct and incorrect responses. These thresholds are arbitrary and should be adjusted given analyst capacity. For this reason, we report performance across all possible levels of analyst intervention.

\subsection{Implementation and Evaluation}

For model evaluation, we utilize GPT-4o, with no additional fine-tuning on vulnerability-specific data. The model processes each software function and generates vulnerability classifications, along with associated token log probabilities. These probabilities are then used to compute the confidence scores, guiding the routing decisions. 

To ensure robustness, we evaluate our methodology under different model settings, analyzing how varying confidence thresholds impact the trade-off between automation and human intervention. Due to the heavy class imbalance in our dataset, we measure the effectiveness of our approach through F1-macro, giving equal weight to both classes. We also provide accuracy as a more intuitive measure of performance although not as informative for this specific task. Lastly, we provide additional analysis of certainty-based routing influencing analyst workload and vulnerability detection performance.

\section{Results}

We break our simulation results down into the three evaluated intelligence levels and corresponding prompting strategies: low-intelligence zero-shot, medium-intelligence few-shot cross-domain, and high-intelligence few-shot in-domain prompts. Table \ref{results} shows a summary of the performance of all routing strategies and intelligence-driven prompt designs at various levels of expert in the loop intervention.

\begin{table}[]
\begin{tabular}{|l|llllllllll|}
\hline
 & \multicolumn{10}{c|}{F1   Score With Different Routing Strategies and EITL Proportions} \\ \hline
 & \multicolumn{2}{c|}{0\%} & \multicolumn{2}{c|}{10\%} & \multicolumn{2}{c|}{25\%} & \multicolumn{2}{c|}{50\%} & \multicolumn{2}{c|}{75\%} \\ \hline
 & \multicolumn{1}{l|}{Rand} & \multicolumn{1}{l|}{UQ} & \multicolumn{1}{l|}{Rand} & \multicolumn{1}{l|}{UQ} & \multicolumn{1}{l|}{Rand} & \multicolumn{1}{l|}{UQ} & \multicolumn{1}{l|}{Rand} & \multicolumn{1}{l|}{UQ} & \multicolumn{1}{l|}{Rand} & UQ \\ \hline
Zero-Shot & \multicolumn{1}{l|}{0.151} & \multicolumn{1}{l|}{0.151} & \multicolumn{1}{l|}{0.193} & \multicolumn{1}{l|}{0.193} & \multicolumn{1}{l|}{0.244} & \multicolumn{1}{l|}{0.266} & \multicolumn{1}{l|}{0.4} & \multicolumn{1}{l|}{0.399} & \multicolumn{1}{l|}{0.64} & 0.601 \\ \hline
FS Cross-Domain & \multicolumn{1}{l|}{0.183} & \multicolumn{1}{l|}{0.183} & \multicolumn{1}{l|}{0.273} & \multicolumn{1}{l|}{0.291} & \multicolumn{1}{l|}{0.372} & \multicolumn{1}{l|}{0.457} & \multicolumn{1}{l|}{0.588} & \multicolumn{1}{l|}{\textbf{0.771}} & \multicolumn{1}{l|}{0.815} & \textbf{0.935} \\ \hline
FS In-Domain & \multicolumn{1}{l|}{0.206} & \multicolumn{1}{l|}{0.206} & \multicolumn{1}{l|}{0.292} & \multicolumn{1}{l|}{0.364} & \multicolumn{1}{l|}{0.484} & \multicolumn{1}{l|}{\textbf{0.55}} & \multicolumn{1}{l|}{0.581} & \multicolumn{1}{l|}{\textbf{0.761}} & \multicolumn{1}{l|}{0.851} & \textbf{0.926} \\ \hline
\end{tabular}
\caption{Illustration of F1 score variation as a function of routing strategy and proportion of data sampled by an Expert-in-the-Loop (EITL). The UQ condition employs the proposed uncertainty quantification technique, whereas Random reflects uniform random sampling without replacement.}
\label{results}
\end{table}

\subsection{Low-Intelligence Zero-Shot Prompting}
\begin{figure*}[h!]
    \centering
    \includegraphics[scale=0.75]{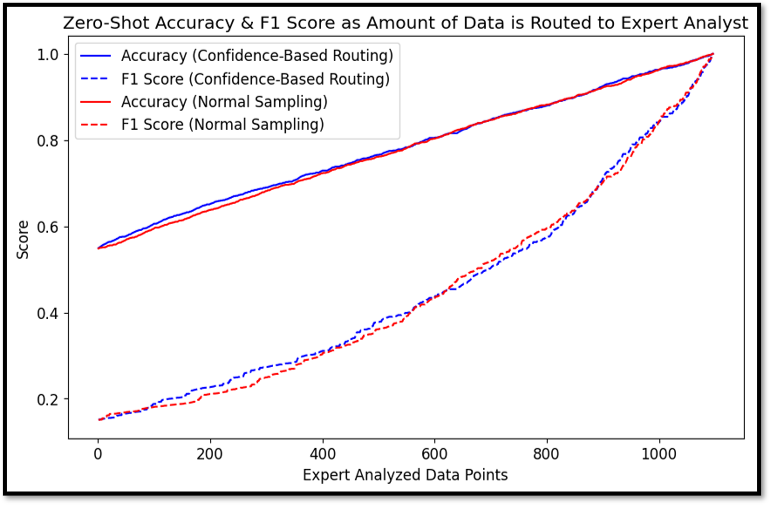}
    \caption{Zero-Shot Accuracy and F1 progression with expert in the loop analysis on varying ranges of data.}
    \label{zero_perf}
\end{figure*}

For our low-intelligence, zero-shot prompting strategy, both initial accuracy and F1 score are poor. Furthermore, confidence-based routing and random sampling by the expert or incident response team proves to be similar in effectiveness. These results indicate that the model is doing little better than randomly guessing if software is vulnerable to exploitation or safe.  
Figure \ref{zero_perf} presents the performance trends across all levels of expert intervention.

\subsection{Medium-Intelligence Few-Shot Cross-Domain Prompting}
For our medium intelligence scenario with few-shot cross-domain prompting, we see a significant improvement in performance. While the initial F1 score remains poor, it improves at a drastically accelerated rate alongside high accuracy. The confidence-based routing strategy also proves to be highly effective, with noticeable improvements in F1 as early as approximately 25\% of the data being routed to an incident response team. This trend continues until normal sampling and confidence-based routing converge upon full dataset evaluation.

These results highlight the ability of LLMs to generalize reported CWEs to identify vulnerabilities and exploits that are potentially unseen. Unlike the zero-shot setting, where the model performed no better than random guessing, few-shot cross-domain prompting enables in-context learning, significantly boosting performance with minimal labeled data. The ability to leverage a small set of examples for generalization across domains is particularly valuable in security-sensitive contexts where annotated data may be scarce.

\begin{figure*}[h!]
\centering 
\includegraphics[scale=0.75]{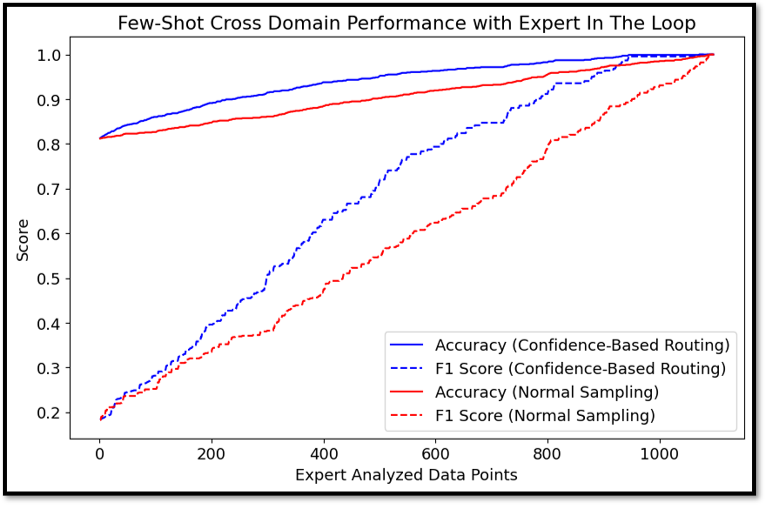} 
\caption{Few-Shot Cross-Domain Accuracy and F1 progression with expert-in-the-loop analysis on varying ranges of data.}
\label{few_cd}
\end{figure*}

Moreover, confidence-based routing ensures that human analysts focus their efforts where they are most needed, maximizing efficiency while maintaining high detection accuracy. By reducing the number of cases requiring manual review, this approach enhances the scalability of vulnerability detection workflows, allowing incident response teams to prioritize high-risk cases without being overwhelmed by the full dataset.

vention thresholds and corresponding F1 and accuracy scores for cross-domain few-shot prompting can be seen in Figure \ref{few_cd}.

\subsection{High-Intelligence Few-Shot In-Domain Prompting}

In our high-intelligence scenario using a few-shot in-domain prompting strategy, we observe that while the model begins with a higher classification performance, there is little variation in F1 scores across different intervention thresholds. Similar to the few-shot cross-domain approach, few-shot in-domain prompting significantly outperforms zero-shot vulnerability detection, particularly at lower levels of expert intervention. However, its performance remains comparable to the cross-domain few-shot approach, suggesting that the model derives limited additional benefit from exact matches within the training examples. Our high-intelligence scenario limits CWEs to MITRE's top 25 most dangerous software weaknesses list, resulting in a reduction of the overall data to 606 data points.

This result may indicate that GPT’s ability to generalize vulnerability patterns is sufficiently strong when provided with relevant examples—regardless of whether they are drawn from the same domain. The similarity between known CWES may allow the model to extrapolate effectively, as long as it receives minimal structured guidance. This suggests that, beyond a certain threshold, exposure to domain-specific vulnerabilities may not provide a significant advantage over cross-domain learning, a stark contrast from traditional fine-tuned models.

\begin{figure*}[h!] 
\centering 
\includegraphics[scale=0.65]{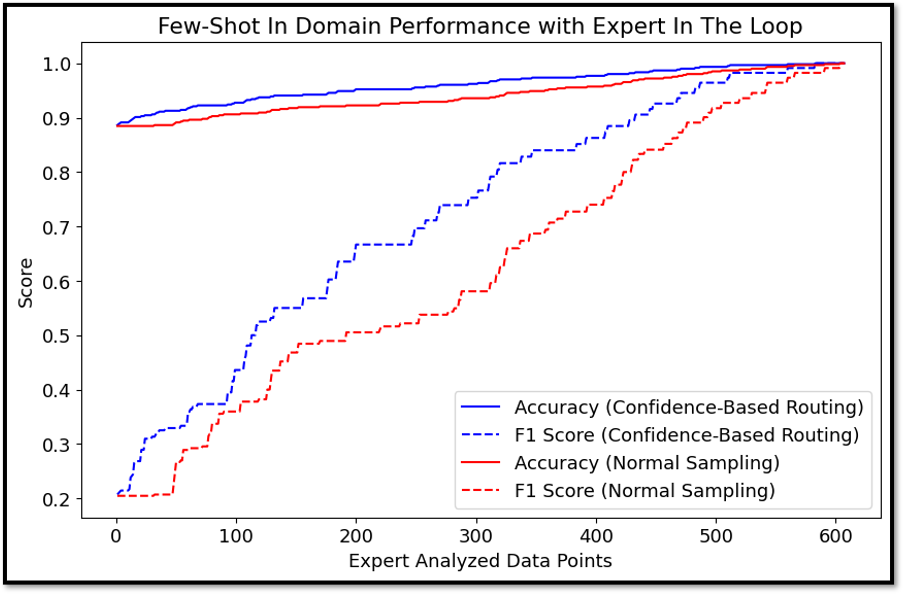}
\caption{Few-Shot In-Domain Accuracy and F1 progression with expert-in-the-loop analysis on varying ranges of data.} 
\label{few_id} 
\end{figure*}

Figure \ref{few_id} illustrates the progression of accuracy and F1 scores under expert-in-the-loop intervention at varying levels of dataset coverage.

\section{Discussion}

Our results demonstrate that leveraging large language models (LLMs) such as GPT-4o for vulnerability detection can provide meaningful improvements in classification performance, particularly when expert intervention is strategically integrated. However, the effectiveness of these models is highly dependent on the prompting strategy and the availability of even minimal examples for guidance.

The poor performance of the zero-shot prompting approach underscores the limitations of LLMs when tasked with domain-specific classification without context. Despite the model's general world knowledge, it struggles to distinguish between vulnerable and safe software in the absence of relevant examples. This suggests that domain adaptation is essential, even in cases where LLMs exhibit strong language understanding. The lack of meaningful performance differences between confidence-based routing and random expert sampling at the zero-shot level further supports this claim, indicating that when a model is uncertain in a fundamentally ill-informed manner, no routing strategy can meaningfully compensate.

The few-shot cross-domain approach significantly improves performance, demonstrating the LLM's ability to generalize across related but distinct vulnerability datasets. While initial F1 scores remain suboptimal, confidence-based routing enables appropriate identification of uncertain cases, allowing expert oversight to be applied where it has the greatest impact. Notably, improvements in F1 become evident with as little as 25\% expert-labeled intervention, underscoring the efficiency of this strategy in real-world applications where expert resources are limited.

The few-shot, in-domain prompting approach yields strong performance but does not provide significant gains over the cross-domain approach. This suggests that the model’s ability to recognize patterns in vulnerability identification is more dependent on structured guidance and close proxies to the task domain, than on exact task domain specificity. In other words, the LLM can extrapolate from related vulnerabilities with minimal loss in performance, provided it has a sufficiently informative prompt. This insight has broad implications for deployment strategies: organizations seeking to integrate AI-assisted vulnerability detection may not require large domain-specific datasets. Rather, they can leverage well-curated cross-domain examples to achieve similar results.

Across both few-shot prompting strategies, confidence-based routing proves to be an effective mechanism for prioritizing expert review. By focusing human oversight on the most ambiguous cases, this approach maximizes the efficiency of security teams, reducing overall workload. The results suggest that in scenarios where resources are constrained, combining LLM classification with confidence-based triage can optimize vulnerability detection workflows.

These findings highlight the potential of LLMs in augmenting cybersecurity workflows, particularly in environments where expert availability is limited. However, they also raise important questions about model reliability, adversarial robustness, and the interpretability of confidence scores. Future work should explore methods for improving model calibration, integrating external knowledge sources, and fine-tuning models on vulnerability-specific datasets. Additionally, researchers can look at other coding languages and more complex constructions of code beyond functions. Finally, further research into human-AI collaboration frameworks can help refine expert-in-the-loop strategies, ensuring that AI systems serve as force multipliers rather than mere decision-support tools.

\section{Conclusion}

This study demonstrates the potential of large language models (LLMs) as powerful tools for vulnerability detection, particularly when paired with expert-in-the-loop strategies. While zero-shot prompting proves ineffective, few-shot prompting with cross-domain settings enables models to generalize with minimal supervision. Importantly, confidence-based routing significantly enhances efficiency, allowing human analysts to focus on high-uncertainty cases where their expertise is most needed.

Our findings suggest that AI-driven vulnerability assessment can complement traditional cybersecurity workflows, reducing response times and improving threat detection at scale. The ability of LLMs to generalize across domains with limited examples underscores their adaptability, offering a path forward for organizations seeking to deploy AI-enhanced security tools without extensive domain-specific fine-tuning.

However, challenges remain. Ensuring model reliability, interpretability, and robustness against adversarial inputs are critical next steps for future research. Additionally, refining human-AI collaboration strategies will be key to maximizing the benefits of confidence-based triage systems. Moving forward, integrating external knowledge sources, fine-tuning models on targeted security datasets, and further optimizing routing strategies can enhance both precision and operational efficiency.

By advancing AI-driven approaches to vulnerability detection, this research provides a foundation for scalable, adaptable, and resource-efficient cybersecurity solutions. As AI capabilities continue to evolve, strategic human-AI partnerships will be essential in maintaining an edge against emerging threats.

\section{Acknowledgements and Disclaimer}
We would also like to feedback received from colleagues Nico Manzonelli, Samuel Ryan, and Eva Brown. The views of this paper are solely those of the authors and not representative of any affiliated organizations.

\footnotesize

\bibliographystyle{wsc}

\bibliography{references}

\section*{AUTHOR BIOGRAPHIES}
\noindent {\bf \MakeUppercase{David Farr}} is a PhD student at the University of Washington in the School of Information Science. He received his MSc in Operational Research with Data Science from the University of Edinburgh, and he did his Bachelor's degree in Systems Engineering from the United States Military Academy at West Point. His research interests include multi-agent systems, network analysis, and data annotation. His e-mail address is \email{dtfarr@uw.edu} and his website is \url{https://davidthfarr.github.io/}. \\

\noindent {\bf \MakeUppercase{Kevin Talty}} is a Cyber Operations Officer, holding the rank of Captain, in the United States Army. Kevin Received a B.S. in Mathematical Sciences from the United States Military Academy at West Point and a Masters in Business Analytics from the Massachusetts Institute of Technology. His interests include Linear/Non-Linear Optimization and Adversarial Machine Learning. His email is \email{kevin.f.talty.mil@army.mil} \\

\noindent {\bf \MakeUppercase{John Stockdale}} is a Cyber Operations Officer, holding the rank of Captain, in the United States Army. John Received a B.S. in Operations Research from the United States Military Academy at West Point and a Masters in Business Analytics from the Massachusetts Institute of Technology. His interests include Linear/Non-Linear Optimization, Machine Learning, and Artificial Intelligence. His email is \email{john.m.stockdale2.mil@army.mil}. \\

\noindent {\bf \MakeUppercase{Alexandra Farr}} is a Technical Program Manager in Microsoft's sovereign cloud team. She received her MSc in Data Science from the University of Edinburgh, and completed her Bachelor's degree in Mathematics from the United States Military Academy at West Point. Her research interests include cloud architecture and generative AI applications to cybersecurity. Her email is \email{afarr@microsoft.com}.\\

\noindent {\bf \MakeUppercase{Iain J. Cruickshank}} is an adjunct faculty member in Carnegie-Mellon University's Department of Software and Societal Systems Department and at Johns Hopkins University. His professional work focuses on the use of multi-modal data in understanding social phenomenon, developing high performing ML systems with no labeled data, and the creation of AI work roles with the government. with His e-mail address is \email{icruicks@andrew.cmu.edu}.\\

\noindent {\bf \MakeUppercase{Jevin West}} is the co-founder of the new Center for an Informed Public at UW aimed at resisting strategic misinformation, promoting an informed society and strengthening democratic discourse. His research and teaching focus on the impact of data and technology on science and society, with a focus on slowing the spread of misinformation. His e-mail address is \email{ jevinw@uw.edu } and his website is \url{https://www.jevinwest.org/}.\\

\vspace{22pt}

\end{document}